# Free carrier absorption in quantum cascade structures

F. Carosella[1], C. Ndebeka-Bandou[1], R. Ferreira[1], E.Dupont[2], K. Unterrainer[3], G. Strasser[4], A.Wacker[5] and G. Bastard[1]

[1] *Laboratoire Pierre Aigrain, Ecole Normale Supérieure, CNRS (UMR 8551), Université P. et M. Curie, Université D. Diderot, 24 rue Lhomond F-75005 Paris, France*

[2] *Institute for Microstructural Sciences, National Research Council, Ottawa, Ontario, Canada K1A0R6*

[3] *Technical University Vienna - Photonics Institute, Gusshausstr. 25-29,  A-1040 Vienna, Austria*

[4] *Technical University Vienna - Solid State Electronics Institute, Floragasse 7, A-1040 Vienna, Austria*

[5] *Mathematical Physics, Lund University, Box 118, S-22100 Lund, Sweden*

**Abstract**. We show that the free carrier absorption in Quantum Cascade Lasers is very small and radically different from the classical Drude result on account of the orthogonality between the direction of the carrier free motion and the electric field of the laser emission.  A quantum mechanical calculation of the free carrier absorption and inter-subband oblique absorption induced by interface defects, coulombic impurities and optical phonon absorption/emission is presented for QCL's with a double quantum well design.  The interaction between the electrons and the optical phonons dominates at room temperature.

Pacs: 73.21.Ac, 78.67.Pt

## I. INTRODUCTION

Quantum Cascade Lasers (QCL) are unipolar structures where the lasing action takes place between the conduction subbands of biased multi quantum well structures [1]. So far, the THz QCL operates only in a limited temperature range and the search for improved structures is worldwide pursued [2,3]. Among possible reasons for the degradation of performances are the depopulation of the upper levels (non radiative escape) but also the re-absorption of the laser photons. The latter are unavoidable because of the free carriers, in particular those that occupy the upper subband of the lasing transition. The free carrier absorption (FCA) is well documented in bulk material where a quantum mechanical calculation [4] leads to a free carrier absorption coefficient that resembles much the semi – classical Drude result [5]. Extrapolating the Drude model from bulk materials to THz QCL leads to a free carrier absorption of the order of $10^2 \, cm^{-1}$, i.e. comparable or larger than actual QCL gains at 2 THz (see e.g [6]). Such large free carrier absorption would jeopardize the future use of QCL. It was however shown [7] in mid infrared QCL that the free carrier absorption plays a small role in the actual laser losses. A calculation [8] of FCA induced by interface roughness in single quantum wells does predict a small absorption coefficient. Another similar calculation of FCA in quasi-2D systems has been realised in presence of acoustical phonons [9].

It is important to stress that there exists a conceptual difficulty inherent to the wave propagation direction when discussing the free carrier absorption in actual cascade structures. The electric vector of the light emitted by QCL is directed along the growth direction, that we will take here parallel to the z axis. The electron states are quasi 2D with bound states along z and extended states along the x and y directions. It results from this configuration that the widely used Drude model to handle free carrier absorption is genuinely inapplicable to QCL (FIG.1) since the carrier free motion occurs in a plane perpendicular to the electric field, thereby making it impossible to rely on a –e**E** term in the Newton law (classical description) or to the existence of intra – subband transitions driven by the electric field (quantum mechanical approach). This remark immediately implies that the often used scaling of the FCA coefficient $\alpha(\omega) \sim \omega^{-p}$ with p ~ 2-3 valid in bulk materials is highly questionable when applied to QCL structures [7]. Along the same line, it may be foreseen that the FCA will be substantially weaker than previously anticipated because the carrier in – plane acceleration by the electric field will be still possible but only because of couplings to the neighboring subbands. As recently shown in [10] the bulk free carrier absorption in superlattices evolves from these inter-subband transitions, and thus all relevant effects are included in a proper treatment of inter-subband processes, as presented below.

FCA is intrinsically connected to scattering and thus a quantitative description of scattering processes is of utmost importance for a quantitative description. Within a density matrix approach, Willenberg *et al*. [11] showed that the optical transition actually take place between two states, with an energy difference equal to the photon energy ℏω. The difference between ℏω and the subband spacing is compensated by a change in the kinetic energy in the in-plane direction as provided by a scattering process accompanying the transition between the bands. In this paper, these scattering assisted transitions are evaluated in detail in a model QCL structure which is similar to the double quantum well design whose lasing action was first demonstrated by Kumar *et al*. [2]. In particular, we focus on transitions, where the scattering process brings the electron back to its original subband. Such intraband scattering-assisted absorption processes are in full agreement with the typical description of FCA in the bulk and are expected to be of particular importance at low frequencies, when other subbands can energetically not be reached. We shall prove that the free carrier absorption in such a double quantum well structure is very small (of the order of 0.1 - 1 cm$^{-1}$). In addition, we shall show that the FCA has a peak in the vicinity of the lowest lying transition energy, that promotes an electron from the upper state of the lasing transition to the nearest subband, while it does not display the $\omega^{-p}$ bulk behaviour characteristic of a Drude - like approach.

## II. MODEL OF FCA

We consider a simplified cascade structure. It comprises $N_p$ periods with thickness $L_z$. The electronic states from each period are taken as independent from those of the adjacent ones. Each period contains an asymmetric double quantum well (DQW) structure made of two GaAs wells ($L_1$ = 23.2nm and $L_2$ = 9.8nm respectively) separated by an intermediate $Ga_{0.85}Al_{0.15}As$ barrier ($L_b$ = 3.1 nm). We neglect the bias electric field to the extent that it does not modify strongly the energy levels and wavefunctions inside a given period. The DQW supports six bound states for the z motion $E_n$, n = 1,2,…, 6. We suppose that the lasing action takes place between $E_2$ and $E_1$ ($\hbar\omega_{21}$ = 16.6 meV). The DQW contains relatively few carriers with an areal concentration equal to $n_e$ = 2.17x10$^{10}$cm$^{-2}$. In the following we will refer to the eigenstates and eigenenergies of a perfect DQW as $\langle \vec{\rho}, z | n, \vec{k} \rangle = \chi_n(z) \frac{1}{\sqrt{S}} e^{i\vec{k}\cdot\vec{\rho}}$ and $\varepsilon_{nk}$ = $E_n$ + $\varepsilon_k$, where $(\vec{\rho}, z) = \vec{r}$, $\varepsilon_k = \frac{\hbar^2 k^2}{2m^*}$ with $\vec{\rho}$ the in plane position and $\vec{k}$ a 2D wave vector.

We are interested in studying the transitions that an electron belonging to the upper level of the laser transition can make because of photon re-absorption. The coupling between the electrons and the electromagnetic wave is provided by the $\vec{A}\cdot\vec{p}$ term where $\vec{A}$ is the vector potential of the wave. Without defects or phonons an electromagnetic wave polarized along z (i. e. propagating in the layer plane, as it is the case in QCL) cannot induce any $|2\vec{k}\rangle \rightarrow |2\vec{k}'\rangle$ transition inside the $E_2$ subband because $\langle 2 | p_z | 2 \rangle = 0$ and, in addition, because the transitions must be vertical in the k space on account of the translation invariance. Thus, we can call "doubly forbidden" the intra-subband transitions in a perfect QCL. Defects, phonons break the in-plane translation invariance but, still, the intra – subband transitions with defects perturbed eigenstates of a given subband remain forbidden because there is no average velocity for a bound state. With our formulation we show that one needs to allow at least for one virtual intermediate coupling in excited subbands to get a non zero intra – subband absorption. In the following, we take $E_3$ to be this excited subband but it is clear that one has to sum over all possible subbands to get a full account of the FCA in actual QCL structures. The computation of the transition rate then becomes very similar to the one of oblique (i.e. $\Delta\vec{k} \neq 0$) inter-band transitions in bulk materials [4, 12]. For comparison, we shall also give results for the inter-subband $|2\vec{k}\rangle \rightarrow |3\vec{k}'\rangle$ transitions, which, unlike the intra-subband case, do not suffer from a vanishing $\langle i | p_z | f \rangle$ matrix element between initial and final subbands. Notice that in the following we will refer only to the *intra*–subband transitions as FCA, because of the analogy between this absorption process and the original FCA in bulk systems.

For an electromagnetic wave with angular frequency ω the energy loss rate $P_{ij}(\omega)$ in presence of static disorder associated with the transitions $|i\vec{k}\rangle \rightarrow |j\vec{k}'\rangle$ is given by

$$P_{ij}(\omega) = \frac{\pi e^2 E_{em}^2}{m^{*2}\omega} \sum_{\vec{k},\vec{k}'} \left(f_{i\vec{k}} - f_{j\vec{k}'}\right)\left|\langle \psi_{i\vec{k}} | p_z | \psi_{j\vec{k}'} \rangle\right|^2 \delta\left(\varepsilon_{j\vec{k}'} - \varepsilon_{i\vec{k}} - \hbar\omega\right) \quad (1)$$

where $\psi_{i\mathbf{k}}$, $\psi_{j\mathbf{k}'}$ are the wavefunctions for the initial and final states taking into account the defects at the first order and $f_{i\mathbf{k}}$, $f_{j\mathbf{k}'}$ are the occupation functions respectively of the initial and final electronic states. In the following, we shall take the occupation functions to be a Boltzmann distribution characterized by an electronic temperature T. Note that the first order correction to the eigenenergies due to defect potentials vanishes. Manipulating the matrix element in (1) we see clearly that, at the lowest order in the defect potential, the transition rate results from the quantum interference of two paths where either the defect potential acts first and the coupling to light follows or vice versa (see FIG.2). Notice that the first order expansion of the perturbed wavefunctions displays energy denominators, which, as shown below, leads to divergences in the absorption coefficient. In a more complete theory where the defects would be considered to all orders in perturbation these divergences would be suppressed and replaced by finite maxima. Qualitatively, replacing $\omega$ by $\omega - i/\tau$ in the transition amplitudes, will have the same effect as resumming all the perturbation series. This implies that the formulae derived below are reliable when $|\omega - \omega_0| \geq 1/\tau$ where $\omega_0$ is the resonant frequency and $\tau$ a typical relaxation time.

To the extent that the laser mode is uniform over the $N_p$ periods of the cascade structure, the absorption coefficient is related to the energy loss rate by

$$\alpha(\omega) = \frac{N_p P(\omega)}{IV} \quad (2)$$

where $I = \varepsilon_0 c n E_{em}^2 / 2$ is the intensity of the incident radiation, $V = N_p L_z S$ ($\varepsilon_0$ is the vacuum dielectric constant, $E_{em}$ the electric field, $c$ the light velocity and $n$ the refraction index). $\alpha(\omega)$ is therefore independent of $N_p$.

In the following, we discuss free carrier absorption mediated by various scattering mechanisms. We shall retain two kinds of static defects: coulombic scatterers and interface defects that have been shown to give rise to level lifetime of a few ps [13]. For completeness, we also investigate the effect of the Fröhlich coupling between electrons and Longitudinal Optical (LO) phonons on the FCA.

## II. 1 INTERFACE DISORDER

The interface defects [13-16] are taken as one monolayer deep protrusions of either the GaAs well into the $Ga_{0.85}Al_{0.15}As$ barrier (attractive defects) or vice versa (repulsive defects). They have a Gaussian shape in the layer plane [16]. For a nominal barrier/well interface located at $z = z_0$ we have:

$$V_{def}(\vec{r}) = V_b g(z) \sum_{\vec{\rho}_j} \exp\left(-\frac{(\vec{\rho} - \vec{\rho}_j)^2}{2\sigma^2}\right) \quad (3)$$

where $g(z)=+Y(z-z_0)Y(h_{def}-z+z_0)$ for repulsive defects and $g(z)=-Y(-z+z_0)Y(h_{def}+z-z_0)$ for attractive defects with the Heaviside function $Y(z)$. $h_{def}$ is the defect height that here we take equal to one monolayer (2.83Å in GaAs), $V_b$ is the potential barrier height. Besides the characteristic in–plane size $\sigma$, the defects are characterized by their areal concentration $n_{def} = N_{def}/S$ or equivalently by the fractional coverage of the surface $fr = \pi\sigma^2 n_{def}$.

For the gaussian interface defects and after averaging over the position of the defects in the layer plane, one obtains after some calculations

$$\alpha_{ij}^{def}(\omega) = \frac{\pi e^2 n_e V_b^2 \sigma^4}{\varepsilon_0 c n m^* L_z \hbar}\left(1 - e^{-\beta\hbar\omega}\right)\frac{|\langle 2|p_z|3\rangle|^2}{\hbar\omega} R_{ij}(\omega) I_{ij}^{def}(\omega) \quad (4a)$$

where $\beta = (k_B T)^{-1}$, $R_{ij}$ is a "resonant factor" respectively for intra-subband (i = j) and inter-subband (i ≠ j) transitions given by

$$R_{22}(\omega) = \left(\frac{1}{\hbar\omega - E_3 + E_2} + \frac{1}{\hbar\omega + E_3 - E_2}\right)^2$$

$$R_{23}(\omega) = \left(\frac{1}{\hbar\omega - E_3 + E_2}\right)^2 \quad (4b)$$

and

$$I_{ij}^{def}(\omega) = F_{ij}^{def} e^{\frac{-2m^*(\hbar\omega - E_j + E_i)\sigma^2}{\hbar^2}} 2\pi \int_0^\infty dx e^{-x(1+C)} I_0\left(C\sqrt{x^2 + \beta x(\hbar\omega - E_j + E_i)}\right) Y\left(x + \beta(\hbar\omega - E_j + E_i)\right) \quad (4c)$$

where $I_0$ is the Bessel function of order zero with an imaginary argument, $C = 4m^*\sigma^2/(\beta\hbar^2)$, and where

$$F_{22}^{def} = \sum_{z_0} \left( n_{att} \left| \int_{z_0 - h_{def}}^{z_0} \chi_3 \chi_2 dz \right|^2 + n_{rep} \left| \int_{z_0}^{z_0 + h_{def}} \chi_3 \chi_2 dz \right|^2 \right) \quad (4d)$$

$$F_{23}^{def} = \sum_{z_0} \left( n_{att} \left| \int_{z_0 - h_{def}}^{z_0} (\chi_3^2 - \chi_2^2) dz \right|^2 + n_{rep} \left| \int_{z_0}^{z_0 + h_{def}} (\chi_3^2 - \chi_2^2) dz \right|^2 \right)$$

are two factors that account for the values of the wavefunctions associated with the states involved in the virtual coupling, close to the disordered interfaces. $n_{att}$ and $n_{rep}$ are the concentrations of attractive and repulsive interface defects. Both expressions for intra- and inter-subband transitions are proportional to the areal density of electrons but also to the number of scatterers. None of them behaves like a Drude term $\omega^{-p}$. In contrast, both of them diverge when the photon energy approaches the energy of the inter - subband transition $E_3 - E_2$.

## II. 2 IMPURITIES

The impurities are taken into account as coulombic scatterers homogenously distributed on planes located at positions $z_n$ [17]. By using the same formalism as in eq.(1-2), one can derive for impurity absorption the following expression:

$$\alpha_{ij}^{imp}(\omega) = \frac{e^6 n_e n_{imp}}{16\pi\varepsilon_0^3 \varepsilon(0)^2 cnm^* L_z \hbar} (1 - e^{-\beta\hbar\omega}) \frac{|\langle 2|p_z|3\rangle|^2}{\hbar\omega} R_{ij}(\omega) I_{ij}^{imp}(\omega) \quad (5a)$$

where $R_{ij}$ is the resonant factor given in eq.(4b) and where

$$I_{ij}^{imp}(\omega) = \sum_{z_n} \int_0^\infty dx e^{-x} Y\left(x + \beta(\hbar\omega - E_j + E_i)\right) \int_0^{2\pi} d\vartheta \frac{F_{ij}^{imp}(Q_{ij}(x,\vartheta,\omega); z_n)}{Q_{ij}^2(x,\vartheta,\omega)} \quad (5b)$$

with

$$F_{22}^{imp} = \int dz \chi_2(z) \chi_3(z) e^{-Q_{22}|z - z_n|}$$

$$F_{23}^{imp} = \int dz \left(\chi_3^2(z) - \chi_2^2(z)\right) e^{-Q_{23}|z - z_n|} \quad (5c)$$

Note that these two functions depend on the localization/delocalization of the wavefunctions on the structure. In eqs(5b, 5c) there is:

$$Q_{ij}^2(x,\theta,\omega) = \frac{2m^*}{\hbar^2 \beta}\left[2x + \beta(\hbar\omega - E_j + E_i) - 2\cos(\theta)\sqrt{x^2 + x\beta(\hbar\omega - E_j + E_i)}\right] \quad (5d)$$

The absorption coefficient due to impurities is on many respects similar to the one derived in the presence of interface disorder: the dependence on the electron concentration $n_e$ and on the areal impurity density $n_{imp}$ is linear and the frequency dependency is not Drude-like and leads to a divergence when the photon energy $\hbar\omega = E_3 - E_2$. In the above formulation we have used an unscreened Coulomb potential. This approximation represents an upper bound for impurity–induced FCA and inter-subband oblique transitions. It is expected to work the better at elevated temperature. In fact the 2D Debye screening length $q_D^{-1} = n_e e^2 / (k_B T \varepsilon_0 \varepsilon(0)) \approx 79 nm$ at T= 300K with $\varepsilon(0)$ = 12.4 the static dielectric constant for GaAs. The oblique virtual transitions are characterized by matrix elements $\langle 2\vec{k}|V_{Coul}|3\vec{k}'\rangle$ or $\langle 3\vec{k}|V_{Coul}|2\vec{k}'\rangle$. Screening of the coulombic potential can be neglected if the wavevector change $\Delta\vec{k} = |\vec{k}'-\vec{k}| \gg q_D$ since at large wavevector transfer the screened and unscreened potentials nearly coincide. But we know that, for example in the case of intra-subband transitions, $k' = \sqrt{k^2 + 2m^*\omega/\hbar}$. Hence, screening can be neglected if $\sqrt{k^2 + 2m^*\omega/\hbar} - k \gg q_D$. We note that $\sqrt{2m^*\omega/\hbar} \approx 10^8 cm^{-1}$ which is typically 9 times larger than $q_D$. Hence, for the more populated states $|\Delta\vec{k}| \gg q_D$ and screening effects can be neglected.

## II. 3 LO PHONONS ABSORPTION AND EMISSION

It is well known that the interaction between electrons and LO phonons dominates the high temperature electronic mobility of III-V and II-VI semiconductors. It is then likely that it should also affect FCA in QCL structures. The energy loss rate $P_{ij}(\omega)$ due to the absorption of a LO phonon, associated with the transitions $|i\vec{k}\rangle \rightarrow |j\vec{k}'\rangle$ is given by

$$P_{ij}^{LOabs}(\omega) = \frac{\pi e^2 E_{em}^2}{m^{*2}\omega} \sum_{\vec{k},\vec{k}',\vec{q}} \left[ f_{i\vec{k}}(1-f_{j\vec{k}'})\left|\langle\psi_{i\vec{k},N_q}|p_z|\psi_{j\vec{k}',N_q-1}\rangle\right|^2 - f_{j\vec{k}'}(1-f_{i\vec{k}})\left|\langle\psi_{j\vec{k}',N_q-1}|p_z|\psi_{i\vec{k},N_q}\rangle\right|^2 \right] \delta(\varepsilon_{j\vec{k}'} - \varepsilon_{i\vec{k}} - \hbar\omega - \hbar\omega_{LO})$$

(6a)

Here, the first term refers to the photon absorption bringing an electron from the perturbed mixed electron-LO phonons state $\psi_{i\vec{k},N_q}$ (containing $N_q$ phonons of energy $\hbar\omega_q \approx \hbar\omega_{LO}$) to the state $\psi_{j\vec{k}',N_q-1}$ with the absorption of a LO phonon; the second term refers to the reverse process: a photon emission bringing back an electron from $\psi_{j\vec{k}',N_{q-1}}$ to $\psi_{i\vec{k},N_q}$ with the emission of a LO phonon. The perturbing potential is the Fröhlich coupling and $\vec{q}$ is the 3D phonon wavevector. The energy loss rate $P_{ij}(\omega)$ due to the emission of a LO phonon, associated with the transitions $|i\vec{k}\rangle \rightarrow |j\vec{k}'\rangle$ is given by a similar expression

$$P_{ij}^{LOemi}(\omega) = \frac{\pi e^2 E_{em}^2}{m^{*2}\omega} \sum_{\vec{k},\vec{k}',\vec{q}} \left[ f_{i\vec{k}}(1-f_{j\vec{k}'})\left|\langle\psi_{i\vec{k},N_q}|p_z|\psi_{j\vec{k}',N_q+1}\rangle\right|^2 - f_{j\vec{k}'}(1-f_{i\vec{k}})\left|\langle\psi_{j\vec{k}',N_q+1}|p_z|\psi_{i\vec{k},N_q}\rangle\right|^2 \right] \delta(\varepsilon_{j\vec{k}'} - \varepsilon_{i\vec{k}} - \hbar\omega + \hbar\omega_{LO})$$

(6b)

Sketches in FIG.3 illustrate these processes.

The absorption coefficient for LO phonon absorption, is given by

$$\alpha_{ij}^{LOabs}(\omega) = \frac{e^4 n_e \omega_{LO}}{16\pi\varepsilon_0^2 \varepsilon_p c n m^* L_z} N_{LO} \frac{|\langle 2|p_z|3\rangle|^2}{\hbar\omega} R_{ij}(\omega) \times$$

$$\left[ \left(1 - e^{-\beta\hbar\omega} e^{(\beta_L - \beta)\hbar\omega_{LO}}\right) I_{ijK}^{LOabs}(\omega) + \frac{n_e \hbar^2 \pi}{2m^* k_B T}\left(e^{-\beta\hbar\omega} e^{(\beta_L - \beta)\hbar\omega_{LO}} - e^{-\beta(\hbar\omega + \hbar\omega_{LO})}\right) I_{ij\Xi}^{LOabs}(\omega) \right]$$

(7a)

where $N_{LO}$ is the Bose occupation function for the phonons and $\beta_L = (k_B T_L)^{-1}$ with $T_L$ the lattice temperature. The resonant factor $R_{ij}$ is given in eq.(4b) and the functions $I_{ijQ}^{LOabs}(\omega)$ with Q = K or Ξ are given by

$$I_{ijQ}^{LOabs}(\omega) = \int_0^\infty dx\, e^{-x} Y\left(x + \beta(\hbar\omega + \hbar\omega_{LO} - E_j + E_i)\right) \int_0^{2\pi} d\vartheta \frac{F_{ij}^{LO}(Q_{ij}(x,\vartheta,\omega))}{Q_{ij}(x,\vartheta,\omega)}$$

(7b)

with the functions

$$F_{22}^{LO} = \int dz \int dz'\, \chi_3(z)\chi_3(z')\chi_2(z)\chi_2(z') e^{-Q_{22}|z-z'|}$$

(7c)

$$F_{23}^{LO} = \int dz \int dz' \left( \chi_3^2(z)\chi_3^2(z') + \chi_2^2(z)\chi_2^2(z') - 2\chi_3^2(z)\chi_2^2(z') \right) e^{-Q_{23}|z-z'|}$$

depending on one of the following expressions respectively for $Q_{ij} = K_{ij}$ or $\Xi_{ij}$

$$K_{ij}^2(x,\theta,\omega) = \frac{2m^*}{\hbar^2\beta} \left[ 2x + \beta(\hbar\omega + \hbar\omega_{LO} - E_j + E_i) - 2\cos(\theta)\sqrt{x^2 + x\beta(\hbar\omega + \hbar\omega_{LO} - E_j + E_i)} \right] \quad (7d)$$

$$\Xi_{ij}^2(x,\theta,\omega) = \frac{m^*}{\hbar^2\beta} \left[ 2x + 2\beta(\hbar\omega + \hbar\omega_{LO} - E_j + E_i) - 2\cos(\theta)\sqrt{x^2 + 2x\beta(\hbar\omega + \hbar\omega_{LO} - E_j + E_i)} \right]$$

Notice that here, as for the other perturbing potentials, we find again the $F_{ij}$ functions that account for the localization/delocalization of the wavefunctions on the structure.

The absorption coefficient in the presence of LO phonon emission is given by:

$$\alpha_{ij}^{LOemiss}(\omega) = \frac{e^4 n_e \omega_{LO}}{16\pi\varepsilon_0^2 \varepsilon_p cnm^* L_z} (N_{LO}+1) \frac{|\langle 2|p_z|3\rangle|^2}{\hbar\omega} R_{ij}(\omega) \times$$

$$\left[ \left(1 - e^{-\beta\hbar\omega} e^{(\beta-\beta_L)\hbar\omega_{LO}}\right) I_{ijK}^{LOemiss}(\omega) + \frac{n_e \hbar^2 \pi}{2m^* k_B T} \left( e^{-\beta\hbar\omega} e^{(\beta-\beta_L)\hbar\omega_{LO}} - e^{-\beta(\hbar\omega-\hbar\omega_{LO})} \right) I_{ij\Xi}^{LOemiss}(\omega) \right] \quad (8a)$$

where the resonant factor $R_{ij}$ is given in eq.(4b) and where the functions $I_{ijQ}^{LOemiss}(\omega)$ with $Q = K$ or $\Xi$ are given by:

$$I_{ijQ}^{LOabs}(\omega) = \int_0^\infty dx\, e^{-x} Y\left(x + \beta(\hbar\omega - \hbar\omega_{LO} - E_j + E_i)\right) \int_0^{2\pi} d\vartheta \frac{F_{ij}^{LO}(Q_{ij}(x,\vartheta,\omega))}{Q_{ij}(x,\vartheta,\omega)} \quad (8b)$$

with the same functions $F_{ij}$ as the ones given for phonon absorption (eq.(7c)) but here depending on one of the following expressions respectively for $Q_{ij} = K_{ij}$ or $\Xi_{ij}$

$$K_{ij}^2(x,\theta,\omega) = \frac{2m^*}{\hbar^2\beta} \left[ 2x + \beta(\hbar\omega - \hbar\omega_{LO} - E_j + E_i) - 2\cos(\theta)\sqrt{x^2 + x\beta(\hbar\omega - \hbar\omega_{LO} - E_j + E_i)} \right] \quad (8c)$$

$$\Xi_{ij}^2(x,\theta,\omega) = \frac{m^*}{\hbar^2\beta} \left[ 2x + 2\beta(\hbar\omega - \hbar\omega_{LO} - E_j + E_i) - 2\cos(\theta)\sqrt{x^2 + 2x\beta(\hbar\omega - \hbar\omega_{LO} - E_j + E_i)} \right]$$

Note that the second terms in eq.(7a) and (8a) gives, as expected from the low carrier concentration and the Boltzmann distribution, a negligible contribution compared to the first term, because the second term has a quadratic dependence on the Boltzmann occupation function, while the first one has only a linear dependence. In structures containing more carriers, one should use Fermi – Dirac distributions for thermalized carriers and the Pauli blocking would play a more important role. Another interesting point concerns the sign of the absorption coefficient and thus the possibility of obtaining gain. The absorption coefficient becomes negative only if the reverse process in eq. (6a) or (6b) becomes dominant. As a matter of fact, $\alpha_{ij}^{LOabs}(\omega) \leq 0$ only if the argument of the exponential in the first term of eq.(7a) is positive; this leads to $T \geq T_L(1 + \omega/\omega_{LO})$ which is a condition that can be verified if $T \neq T_L$. On the contrary, $\alpha_{ij}^{LOemi}(\omega) \leq 0$ only if $T \leq T_L(1 - \omega/\omega_{LO})$, which can never happen because $T \geq T_L$.

### III. RESULTS AND DISCUSSION

Calculations are done for a set of DQW structures where we increase simultaneously the width of the two wells by adding a multiple of one monolayer while keeping constant the central barrier width (23.2nm + $ph_{def}$ /3.1nm /9.8nm + $ph_{def}$, p = 0,1,2…). This procedure allows decreasing $E_2 - E_1$ while the distance $E_3 - E_2$ is kept roughly constant (at 6.6 meV) and the matrix element $\langle 3|p_z|2\rangle$ is reduced by a factor of ~2.

However, note that the $|\langle 3|p_z|2\rangle|^2/\hbar\omega$ factor remain roughly constant for all structures. The carrier effective mass has been taken equal to m* = 0.067 m$_0$.

FIGS.4 show the $\hbar\omega = E_2 - E_1$ dependence of the absorption coefficient in presence of defects $\alpha^{def}(\omega)$. Results are given for several electronic temperatures. The fractional coverage by interface defects was kept at *fr* = 30% and the defect size at σ = 10.8 nm. In FIG.4(a) we show plots of the absorption coefficient for the $|2\vec{k}\rangle \to |2\vec{k}'\rangle$ transitions and in FIG.4(b) the one for the $|2\vec{k}\rangle \to |3\vec{k}'\rangle$ for comparison. We see firstly that the FCA is very small (about 10$^{-2}$ cm$^{-1}$ far away from resonance energy) in agreement with [8] but in stark contrast with the extrapolation of Drude results valid for bulk materials. This small value is due to three main causes: a) the small electron concentration present in THz QCL, b) the doubly forbidden nature of intra - subband transitions and c) the fact that the interface defects are relatively mild scatterers. It is worth stressing that the scattering induced inter-subband absorption (FIG. 4(a)) is about one order on magnitude larger than the free carrier - like intra-subband absorption (FIG. 4(b)). The main reason for such a trend is the smaller wavevector transfer in the former case than in the latter, as evidenced in the arguments of the exponential in the $I_{ij}^{def}(\omega)$ function and the I$_0$ function in eq.(4c).

We show on FIG.5(a) the absorption coefficients for $|2\vec{k}\rangle \to |2\vec{k}'\rangle$ and $|2\vec{k}\rangle \to |3\vec{k}'\rangle$ transitions induced by ionized impurities. These impurities lay in the thinnest QW and for numerical purpose we distributed them on n = 20 equidistant planes. Each plane has an impurity density of (2.17/n) x 10$^{10}$cm$^{-2}$. In FIG.5 (b) we show the absorption coefficient for impurity-induced FCA and oblique inter-subband absorption in presence of residual ionized impurities with a typical volume concentration for GaAs of 3x10$^{15}$cm$^{-3}$. The absorption coefficient is several orders of magnitude larger in FIG.5(a) than in FIG.5(b) because in the first case all the impurities are concentrated in a well where the electronic wavefunction is significant.

The curves shown in FIGS.5(a,b) display the same trends as found when the transitions are induced by the interface defects. The magnitude of the absorption in presence of interface defects is quite comparable to the one obtained with residual doping. We note that both FCA and oblique inter-subband absorption decrease with increasing temperature. Although T appears in several places in eqs.(4-5), the main factor that contributes to the decreased absorption at elevated T is (1-exp(-β$\hbar\omega$)). Physically, this term represents the increasing part played by the stimulated emission that decreases the net absorption coefficient of a Boltzmann thermalized population with fixed carrier concentration. Notice that, from a similar argument, if we were to draw the FCA versus ω for a fixed geometry of the QCL, the curves would not be symmetric around the E$_3$-E$_2$ resonance.

We show in FIGS.6(a,b) the free carrier absorption $|2\vec{k}\rangle \to |2\vec{k}'\rangle$ and the oblique inter - subband absorption $|2\vec{k}\rangle \to |3\vec{k}'\rangle$ due to LO phonon absorption versus $\hbar\omega = E_2 - E_1$. The absorption coefficients $\alpha^{LOabs}(\omega)$ are proportional to the LO phonon occupation at the lattice temperature T$_L$. Hence, at low T$_L$, the LO phonon absorption is inefficient, as expected, but starts to be stronger than interface disorder and residual doping around 100 K and approaches 0.1 - 1 cm$^{-1}$ far away from resonance energy at T = 150 K. The curves show also that the difference between the electronic temperature and the lattice temperature has an effect on the magnitude of the absorption coefficient. This is clearly visible in FIG.6 (a) where the absorption coefficient at T$_L$ = 100 K and T = 150 K is negative (while the ones calculated with T = T$_L$ are always positive) and its absolute value decreases steadily with frequency because the occupation factor is dominating the behavior of this curve (while this is not the case when T = T$_L$).

We present the results of the calculations of the FCA associated with the LO phonon emission in FIGS.7(a,b). We see that the absorption coefficient has the same temperature and $\hbar\omega = E_2 - E_1$ dependences for phonon emission and for phonon absorption. Besides, the order of magnitude of these two absorption coefficients is comparable and higher than the one of the absorption coefficient due to the presence of interface disorder or residual doping.

It is not immediate to interpret the ω dependences of the intra and inter-subband absorptions induced by the defects and by the phonons. The main feature common to all our results is the strong increase of α(ω) when $\hbar\omega$ approaches the intersubband transition energy E$_3$-E$_2$ = 6.6 meV. In our formulation this behaviour comes from the "resonance factor" (eq.4b) which diverges when $\hbar\omega$ = 6.6 meV and which appears because we limit the perturbation expansion to the first order (see the discussion in section II). The other feature common to the various absorption coefficients is that the ones corresponding to FCA increase slightly or level off with increasing ω while the ones corresponding to inter-subband oblique transitions decrease steadily with increasing ω. In order to give an explanation of these different ω dependences, we recall that the large (small) ω values correspond to thin (thick) wells. As a result, the eigenstates are more delocalized at large ω than at small ω [17]. Looking at the expressions for the absorption coefficients we notice that the FCA and the inter-subband processes differ at large ω (far from resonance) only by the functions $I_{ij}$. Such functions contain integrals over z that, in the formulation for interface defects and impurities, involves either $\chi_3^2(z) - \chi_2^2(z)$ or $\chi_3(z)\chi_2(z)$. These two functions behave differently versus ω. We expect that increasing ω (thus increasing the delocalisation of the wavefunctions) the $\chi_3^2 - \chi_2^2$ factor decreases for most values of z, because the squares of the wavefunctions compensate each other. On the other hand, the variation with ω of the factor $\chi_2\chi_3$ is more difficult to predict, because it strongly depends on the z position. Calculating its value for different z we found that on average it does not vary much with ω. We notice that the ω dependences of these two factors are the same as the one obtained for the intra-subband and inter-subband absorption coefficient. Thus, we can conclude that the calculated ω dependence of the absorption coefficient is determined by the localization/delocalization of the wavefunctions which varies with the structure employed and, consequently, with the lasing photon energy $\hbar\omega$. A similar discussion could be made for the absorption coefficient due to the electron - LO phonon interaction, because the formulation depends again on similar relations between the wavefunctions (eq. 7c).

### III-1 CONCLUDING REMARK

Before concluding we discuss briefly the differences between our model and Unuma *et al*.'s approach [15] for the calculation of the inter-subband absorption. The computation of defect induced intra-subband and inter-subband transitions (here $|2\vec{k}\rangle \to |2\vec{k}'\rangle$ and $|2\vec{k}\rangle \to |3\vec{k}'\rangle$) produces lineshapes that are different from the tail of the quasi – lorentzian lineshape derived e.g. by Unuma *et al*. for intersubband absorption. This is because the two calculations are performed in quite different limits of validity. In Unuma *et al*., one starts from allowed inter-subband transitions (i. e. vertical in k). Without broadening the absorption coefficient $\alpha_{23}(\omega)$ is a delta function of the argument ($\hbar\omega - E_3 + E_2$). Scattering broadens this delta function into a lorentzian. Let us remark that the integrated absorption coefficient $\int \alpha(\omega)d\omega$ is essentially independent of the defect concentration (just because integrating a normalized lorentzian gives a quantity that does not depend on the broadening parameter of the lorentzian). Hence, in Unuma *et al*.'s type of calculations, one finds $\int \alpha(\omega)d\omega \approx (n_{def})^0$. In our calculations, such a $(n_{def})^0$ term is missing and at the lowest order we find instead an integrated absorption coefficient that is *linear* in $n_{def}$. This is because we focus our attention on the oblique in k absorptions that are forbidden in the absence of defects. As discussed above (section II), our calculation of the FCA is reliable when the photon energy differs from the resonant one by a typical energy broadening. Note that this markedly off resonant condition appears to be what happens for the FCA in actual QCL lasers: usually the lasing photons are not resonant with another inter-subband transition. Besides, the $|2\vec{k}\rangle \to |2\vec{k}'\rangle$ intra-subband absorption is something that cannot exist in Unuma *et al*. derivation since it would involve a zero oscillator strength (cf eq.2 in [15]). Hence, to get such a non-vanishing contribution, it is mandatory to include a virtual coupling to $E_3$, i.e. to consider perturbation of the current operator*)*. It is also interesting to point out that Unuma *et al*.'s approach at large detuning predicts an absorption that varies like $\Gamma(\hbar\omega - E_3 + E_2)^{-2}$. If we look at our expressions for absorption we find different behaviours: not only do we have this term but several other multiplicative factors that are photon energy dependent. But in the large detuning limit and mild scatterers, our modelling should become "exact". A more detailed comparison between Unuma *et al*.'s model and our perturbative approach is beyond the scope of this paper.

### IV. CONCLUSIONS

In conclusion, we have presented a theoretical analysis of the free carrier absorption in THz QCL. We have shown that a quantum mechanical calculation of the intra-subband transitions leads to very small absorption coefficients for the THz laser photon at current operation temperature. Oblique (in k space) inter – subband transitions, if energetically possible, are more efficient agents for re – absorbing the laser photons. We found that interface defects and ionized impurities (residual doping) are both relatively inefficient for the parameters we used and which are adapted to present THz QCL's. Electron - LO phonons dominate the FCA at room temperature.

**Acknowledgements.** We acknowledge very useful discussions with Dr. A. Vasanelli. The work in Paris has been supported by an ANR contract : ROOTS. The work at Vienna has been supported by FWF (SFB "IR-ON" F25). The work at Lund has been supported by the Swedish Research Council (VR). One of us (GB) acknowledges the Wolfgang Pauli Institute (Vienna) for support.

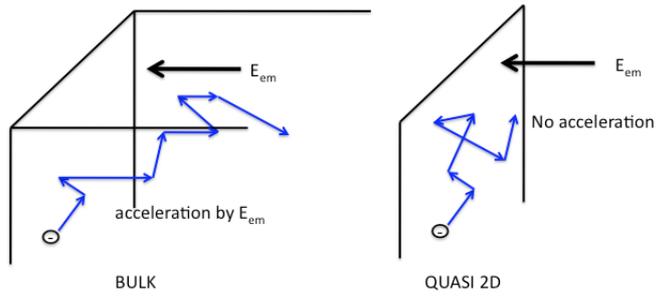

FIG.1 Sketch of the difference between bulk and quasi 2D situations when considering the combined actions of the electric field of an electromagnetic wave and the scatterers on the semi – classical motion of an electron.

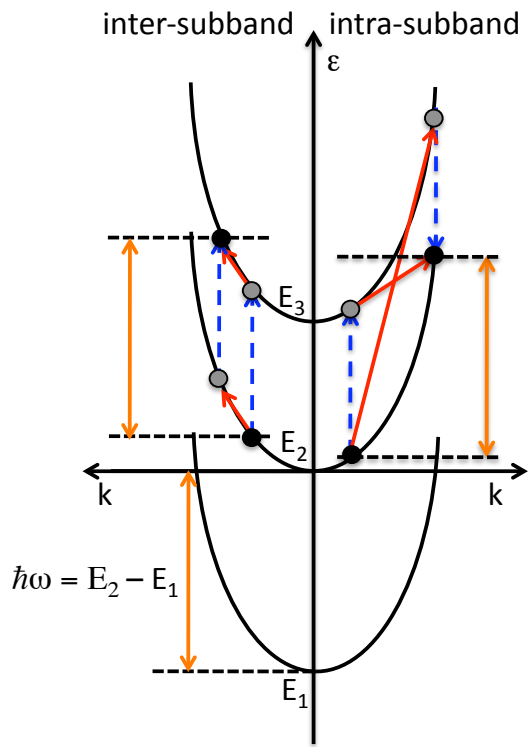

FIG.2 Energy dispersion of the $E_1$, $E_2$ and $E_3$ subbands. Right side: quantum mechanical paths followed by an electron to undertake an intra – subband oblique absorption mediated by static scatterers. Left side: Right panel: quantum mechanical paths followed by an electron to undertake an intra – subband oblique absorption mediated by static scatterers. Left panel: quantum mechanical paths followed by an electron to undertake an inter – subband oblique absorption mediated by static scatterers. Dotted lines refer to electron-photon interaction; full lines refer to electron – defect interaction. Black dots are initial and final states, grey dots are virtual intermediate states.

(a) Intrasubband absorption assisted by LO phonon absorption

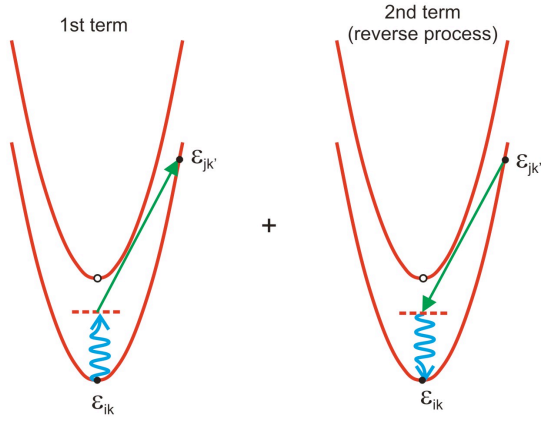

(b) Intrasubband absorption assisted by LO phonon emission

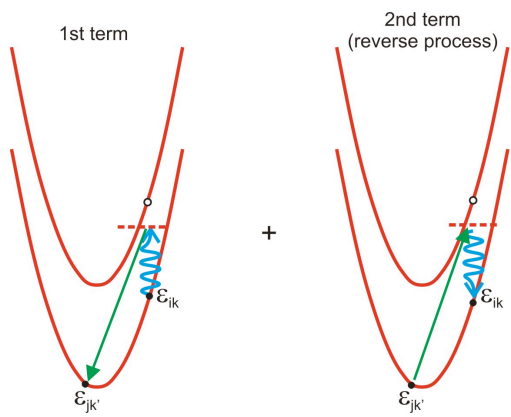

FIG.3 Schematic representation of the electronic intra-subband transition via an intermediate virtual state (represented by a dotted line). Wavy arrows represent transitions due to photons absorption/emission and straight arrows represent transitions due to phonons absorption emission.(a) left panel: photon absorption assisted by one LO-phonon absorption; right panel: photon emission assisted by one LO-phonon emission. (b) left panel: photon absorption assisted by one LO-phonon emission; right panel: photon emission assisted by one LO-phonon absorption.

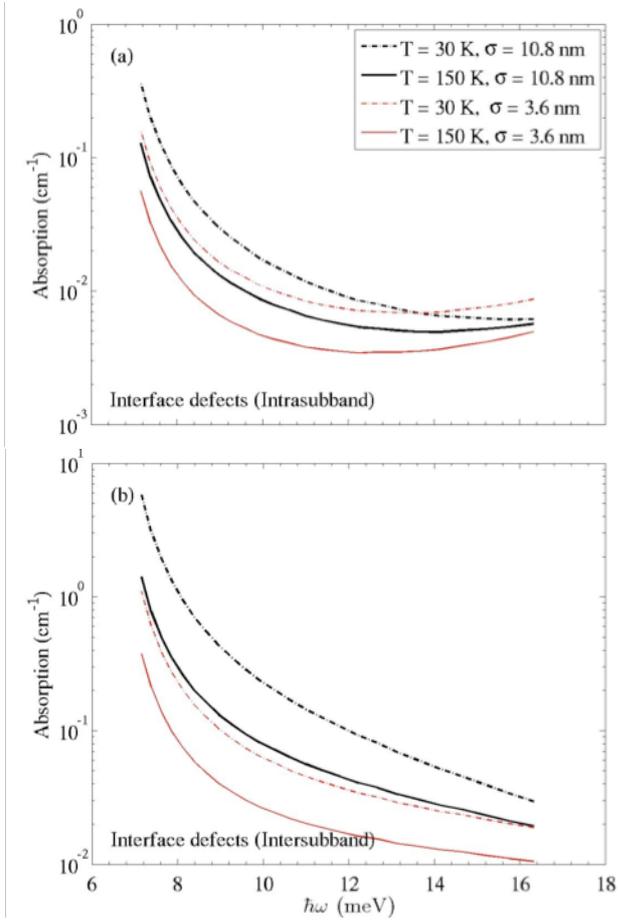

FIG.4 (a) Absorption coefficient $\alpha^{def}(\omega)$ versus $\hbar\omega$ for intra – $E_2$ subband oblique transitions due to interface defects when $\hbar\omega = E_2 - E_1$ is varied (see text) and several electronic temperatures T . (b) Absorption coefficient $\alpha^{def}(\omega)$ versus $\hbar\omega$ for inter - subband $E_2 \to E_3$ oblique transitions due to interface defects when $\hbar\omega = E_2 - E_1$ is varied (see text) and several electronic temperatures T.

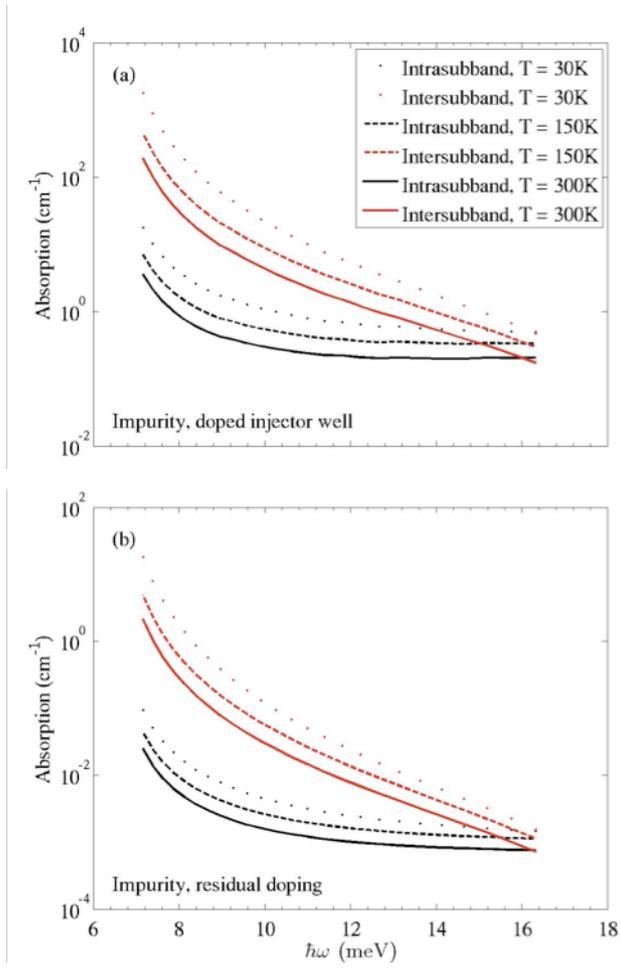

FIG.5 Absorption coefficient $\alpha^{imp}(\omega)$ versus $\hbar\omega$ for intra-subband (black curves) and inter-subband transitions (red curves) due to ionized impurities: (a) doping of the thinnest well, $n_{imp}=2.17 \times 10^{10}$ cm$^{-2}$ (b) residual doping of the whole structure, $n_{imp}=3 \times 10^{15}$ cm$^{-3}$.

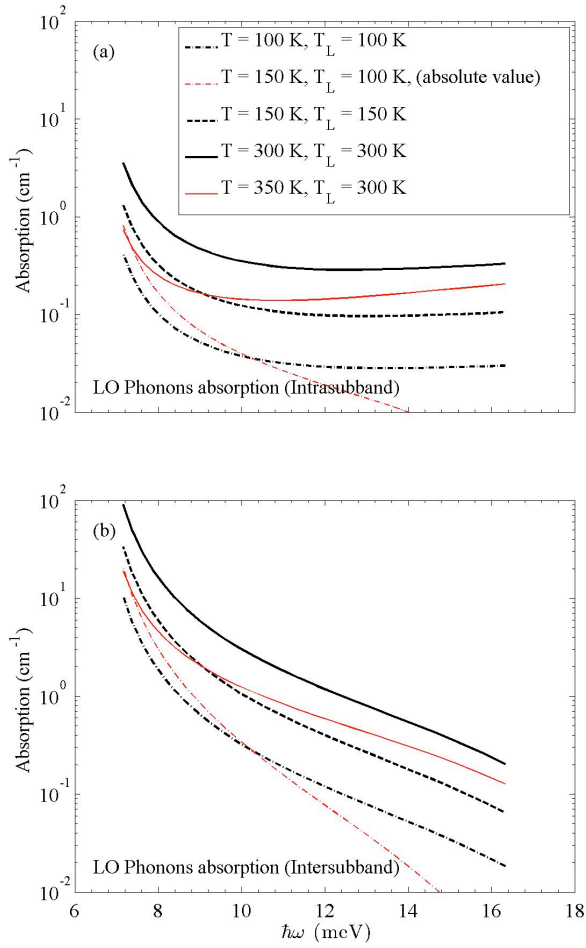

FIG.6(a) Absorption coefficient $\alpha^{LOabs}(\omega)$ versus $\hbar\omega$ for intra - subband transitions due to LO phonon absorption when $\hbar\omega = E_2 - E_1$ is varied (see text). (b) Absorption coefficient $\alpha^{LOabs}(\omega)$ versus $\hbar\omega$ for inter - subband oblique transitions due to LO phonon absorption when $\hbar\omega = E_2 - E_1$ is varied (see text). $T(T_L)$ is the electronic (lattice) temperature.

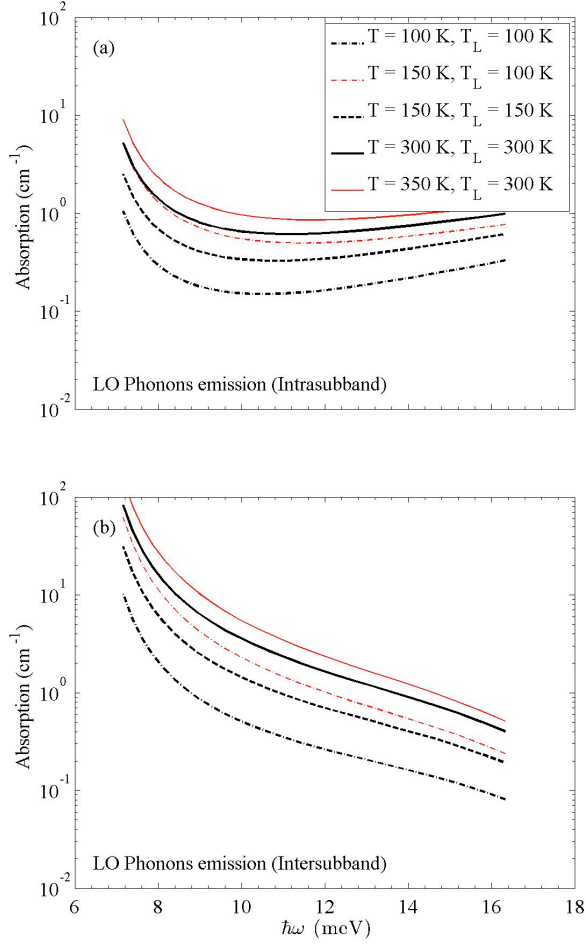

FIG.7(a) Absorption coefficient $\alpha^{LOemi}(\omega)$ versus $\hbar\omega$ for intra - subband transitions due to LO phonon emission when $\hbar\omega = E_2 - E_1$ is varied (see text). (b) Absorption coefficient $\alpha^{LOemi}(\omega)$ versus $\hbar\omega$ for inter - subband oblique transitions due to LO phonon emission when $\hbar\omega = E_2 - E_1$ is varied (see text). $T(T_L)$ is the electronic (lattice) temperature.